\documentclass[twocolumn]{aastex6}
\usepackage{graphicx}
\usepackage[countmax]{subfloat}
\usepackage{booktabs}
\usepackage{mwe}

\begin{document}
\title{X-ray Flux and Spectral Variability of Six TeV Blazars with \textit{NuSTAR}}
\author{Ashwani Pandey\altaffilmark{1,2}, Alok C. Gupta\altaffilmark{1,2}, \& Paul J. Wiita\altaffilmark{3}}

\altaffiltext{1}{Aryabhatta Research Institute of Observational Sciences (ARIES), Manora Peak, Nainital 263002, India; ashwanitapan@gmail.com}
\altaffiltext{2}{Department of Physics, DDU Gorakhpur University, Gorakhpur 273009, India; acgupta30@gmail.com}
\altaffiltext{3}{Department of Physics, The College of New Jersey, 2000 Pennington Rd., Ewing, NJ 08628-0718, USA; wiitap@tcnj.edu}


\begin{abstract}
We report the first results of timing and spectral studies of  \textit{Nuclear Spectroscopic Telescope Array (NuSTAR)} observations of six TeV emitting high-frequency peaked blazars: 1ES 0347$-$121, 1ES 0414$+$009, RGB J0710$+$591, 1ES 1101$-$232, 1ES 1218$+$304 and H 2356-309. Two out of these six TeV blazars, 1ES 1101$-$232 and 1ES 1218$+$304, showed strong evidence of intraday variations in the 3--79 keV energy range during those observations. We also found a hint of an intraday variability timescale of 23.5 ks in the light curve of 1ES 1218$+$304 using an autocorrelation function analysis. We obtained magnetic field $B \sim 0.03$ G, electron Lorentz factor $\gamma \sim 2.16 \times 10^6 $ and emission region size $R \sim 1.19 \times 10^{16} $ cm for 1ES 1218$+$304 using that variability timescale. The other blazars' light curves do not show any variability timescales shorter than their observation lengths; however, we note that the data were both noisier and sparser for them. We also investigated the spectral shape of these TeV blazars and found that 
the spectrum  of 1ES 0414$+$009 is well described by a single power-law with photon index $\Gamma \sim 2.77$. The spectra of the other five HBLs are somewhat better represented by log-parabola models with local photon indices (at 10 keV) $\alpha \sim 2.23-2.67$ and curvature parameters, $\beta \sim 0.27-0.43$.
\end{abstract}


\section{Introduction} \label{sec:intro}

Blazars are the subclass of active galactic nuclei (AGNs) possessing a relativistic jet that is aligned close ($\leq 10^{\circ}$) to the observer's line of sight \citep{1995PASP..107..803U}. The Doppler boosted non-thermal emission from the relativistic jets is highly variable at all observed timescales over essentially the entire electromagnetic spectrum. Variability seen over a timescale of less than a day is called intraday variability (IDV) or microvariability \citep{1995ARA&A..33..163W}, variations over a few days to months are often termed as short term variability (STV) while fluctuations observed over several months to years, or even decades, are known as long term variability (LTV) \citep{2004A&A...422..505G}.  The two classical subclasses of blazars are the BL Lacertae objects (BL Lacs) which have no detectable, or very weak (EW $< 5$\AA), optical emission lines  \citep{1996MNRAS.281..425M} and the flat spectrum radio quasars (FSRQs), which have the usual  strong quasar emission lines in their optical spectra. The two broad bumps seen in the broadband spectral energy distributions (SEDs) of blazars indicate two different  emission mechanisms.  
The low energy peak is well understood to be caused by the synchrotron emission from relativistic electrons in the jet. However, the origin of high energy peak is still under debate. In the leptonic model, the high energy component is interpreted as the inverse Compton (IC) scattering of synchrotron photons themselves (synchrotron-self Compton, SSC; e.g. \citep{1996ApJ...461..657B}) or external photons (external Compton, EC; e.g. \citep{1995ApJ...441...79B}) by the same electrons responsible for the synchrotron emission.  In the alternative hadronic models,  processes such as proton and muon synchrotron emission are thought to  be responsible for the high energy bump (e.g. \cite{2007Ap&SS.307...69B}).  Blazars are also classified through the value of the peak frequency of the synchrotron component.  It typically lies in the infrared to optical region in the low-frequency peak blazars (LBL)  while in the high-frequency peaked blazars (HBL)  it is located at FUV to X-ray energies \citep{1995ApJ...444..567P}. The high energy components of blazar SEDs peak at GeV energies in LBLs and at TeV energies in HBLs, but some LBLs and the intermediate-peaked blazars (IBLs) have still been detected at TeV energies.  

The X-ray emissions of TeV blazars are found to be highly variable at IDV timescales (e.g. \cite{2017ApJ...841..123P} and references therein) as it corresponds to the high-energy tail of the synchrotron component of their SED. 
 Study of X-ray variability at IDV timescales is useful in understanding the underlying physical mechanisms and in constraining the properties of emitting regions. 
The nature of X-ray spectra of TeV emitting blazars have been examined for quite some time. \cite{1990ApJ...360..396W}  found that a single power-law with spectral indices $\sim 1.0$ provided acceptable fits to the X-ray spectra of 24 BL Lac objects observed with the \textit{Einstein Observatory}. The X-ray spectra of a large sample of BL Lac objects in \textit{BeppoSAX} spectral survey of BL Lacs  were well described by either a single power-law or a broken power-law \citep{2002A&A...383..410B}. In more recent studies, the X-ray spectra of TeV blazars were found to be curved at high energies and were better fitted with a log-parabola model (e.g. \cite{2002babs.conf...63G,2005A&A...433.1163D,2007A&A...466..521T,2008A&A...478..395M}).  The log-parabola model was first used by \cite{1986ApJ...308...78L} to better describe the synchrotron emission of BL Lac objects but they didn't provide any physical explanation of the model. Later, \cite{2004A&A...413..489M,2004A&A...422..103M,2006A&A...448..861M} described the X-ray spectra of TeV BL Lac objects Mrk 421 and Mrk 501 in terms of the curved log-parabola model and also gave a possible interpretation of this model in terms of statistical particle acceleration, by assuming that the probability of increase in the energy of emitting particle is a decreasing function of its energy. In a recent study with \textit{Swift/XRT}, \cite{2016MNRAS.458...56W} have found that most of the X-ray spectra  of TeV emitting blazars are well described by the log-parabola model.

The X-ray observatory, \textit{Nuclear Spectroscopic Telescope Array (NuSTAR)}, launched in 2012, consists of two co-aligned hard X-ray telescopes that focus on two almost identical detector modules, Focal Plane Module A (FPMA) and Focal Plane Module B (FPMB)  \citep{2013ApJ...770..103H}. Its high spectral resolution and low background have provided unprecedented sensitivity in the 3--79 keV energy range. 

Until 2005 there were only seven known TeV blazars: Mrk 421, Mrk 501, 1ES 2344+514, PKS 2155--304, 1ES 1959+650, 1ES 1426+428,  
and PKS 2005-489. Thanks to the {\it Fermi} satellite and the ground based TeV gamma-ray facilities (e.g. 
{\it H.E.S.S.} (High Energy Stereoscopic System), {\it MAGIC} (Major Atmospheric Gamma-ray Imaging Cherenkov Telescopes), 
{\it VERITAS} (Very Energetic Radiation Imaging Telescope Array System), and {\it HAWC} (High-Altitude Water Cherenkov Observatory) 
  new TeV blazars have been discovered. In the TeV source catalogue (TeVCat\footnote{\url{http://tevcat.uchicago.edu}}) 
the total number of blazars, at the time of writing, is 64 (HBLs=48, IBLs=8, LBLs=2, FSRQs=6), out of which \textit{NuSTAR} 
has observed only 15 (HBLs = 11, IBLs = 2, FSRQs = 2). The main motivation of this work is to examine the X-ray intraday flux variability  
and the spectral shape of TeV HBLs in the energy range 3--79 keV. 
Blazar variability on IDV timescales 
is one of the most puzzling issues in the field as it requires large energy outputs within small physical scales, and these emission regions  are often very close to the supermassive black hole (SMBH). The blazars we study in the present work are relatively newly listed in the TeV catalogue, and there is essentially no previous
study of these sources in hard X-ray energies. Here we present the first result of X-ray IDV and spectral 
studies of these TeV HBLs in the energy range 3--79 keV.       

In our earlier work \citep{2017ApJ...841..123P} we  examined \textit{NuSTAR} LCs of five TeV HBLs, 1ES 0229$+$200, Mrk 421, Mrk 501, 1ES 1959$+$650 and PKS 2155$-$304, for IDV. 
The \textit{NuSTAR} spectra of these HBLs have been studied by other authors and found to be well described by either the simple power-law  or the curved log-parabola model.   For Mrk 421 see \cite{2015A&A...580A.100S} and \cite{2016ApJ...819..156B}; for Mrk 501 see \cite{2015ApJ...812...65F}; for PKS 2155$-$304 see \cite{2016ApJ...831..142M}; and for 1ES 0229$+$200 and 1ES 1959$+$650 see \cite{2017arXiv171009910B}.
In this work, we present the flux and spectral variability study of the remaining 6 TeV HBLs  observed by \textit{NuSTAR}: 1ES 0347$-$121, 1ES 0414$+$009, RGB J0710$+$591, 1ES 1101$-$232, 1ES 1218$+$304 and H 2356$-$309. We investigate the shapes of the hard X-ray spectra of these TeV HBLs  using both  single power-law models and  log-parabola models. 

The structure of the paper is as follows. The observations and data processing are described in section \ref{sec:data} and the data analysis techniques used to study IDV flux variability and spectral shape of TeV blazars are discussed in section \ref{sec:analysis}. Section \ref{sec:result} presents the results of our study. Sections \ref{sec:discussion} and \ref{conclusion} include a detailed discussion and our conclusions, respectively.

\section{Observations and Data Reduction} \label{sec:data}

We downloaded all \textit{NuSTAR} data sets  that are publicly available from the HEASARC Data archive\footnote{\url{http://heasarc.gsfc.nasa.gov/docs/archive.html}} with good exposure times (those unaffected by passage through the South Atlantic Anomaly (SAA) or other periods of exceptionally high background) greater than 5 ks for these 6 blazars: 1ES 0347$-$121, 1ES 0414$+$009, RGB J0710$+$591,  1ES 1101$-$232, 1ES 1218$+$304 and  H 2356$-$309.   It turns out that there were only single such  observations for each source and they were made between 2015 September 1 and 2016 May 18; the good exposure times ranged from 21.90 to 50.79 ks. Five out of six TeV HBLs were observed with {\it NuSTAR} in different guest observer programs on AGN, but not as targets of opportunity, while H 2356$-$309 was observed in extragalactic surveys performed during the 2.5 years of the {\it NuSTAR} prime mission.
The observing log of the \textit{NuSTAR} data for these six HBL TeV blazars is given in Table \ref{tab:obs_log}.

The \textit{NuSTAR} data were processed using HEASOFT\footnote{\url{http://heasarc.gsfc.nasa.gov/docs/nustar/analysis/}} 
version 6.19 and the updated Calibration Database (CALDB) files version 20161207. The calibration, cleaning and screening of data were done using the standard \textit{nupipeline} script with saamode=OPTIMIZED to correct for SAA passage. Each source LC and spectrum were extracted from a circular region centered at the source using the \textit{nuproducts} script. Background 
data for each source were extracted from circular regions on the same detector module on which the source was focused but free from source contamination. The radii of the source and background regions we used  for the reduction of the data on our six TeV blazars are listed in Table \ref{tab:obs_log};  the brightest source, 1ES 1101$-$232, required a larger extraction radius than the others. 

We summed the background-subtracted count rates of the two nearly identical \textit{NuSTAR} detectors, FPMA and FPMB, and binned them in 5 minute intervals to generate the final light curves (LCs).  The mean values of the difference between count rates between the FPMA and FPMB detectors were only $-$0.004, 0.014, $-$0.015, $-$0.005, $-$0.006 
and $-$0.084 for 1ES 0347$-$121, 1ES 0414$+$009, RGB J0710$+$591, 1ES 1101$-$232, 1ES 1218$+$304, and H 2356$-$309, 
respectively, and as all were essentially constant during the observations a direct sum of the rates is justified.  We used the same 5 minute bins in our earlier work \citet{2017ApJ...841..123P}, and using longer or shorter ones for different objects do not change the IDV LC patterns. The response files (\textit{rmf} and \textit{arf} files) were generated using  \textit{numkrmf} and \textit{numkarf} modules, respectively, within \textit{nuproducts} script.



\begin{table*}
\centering
\caption{Observation log of \textit{NuSTAR} data for six TeV HBLs with radii of source and background regions used}
 \label{tab:obs_log}
 \begin{tabular}{lcclcccc}
  \hline
Blazar Name 	   & ~~Obs. Date  & Start Time (UT) & ~~~Obs. ID 	      & Total Elapsed & Exposure  & Source  & Background  \\
                   & ~~yyyy-mm-dd & hh-mm-ss    &                &      Time (ks)    &  Time (ks)    &radius &   radius \\
\hline
1ES 0347$-$121	&  2015-09-10  &  04:51:08 & 60101036002 &  ~61.27 &  32.93 &$30^{\prime\prime}$  & $30^{\prime\prime}$  \\ 
1ES 0414$+$009  &  2015-11-25  &  17:01:08 & 60101035002 & 107.06 &  34.16 &$30^{\prime\prime}$  & $30^{\prime\prime}$ \\ 
RGB J0710$+$591	&  2015-09-01  &  12:11:08 & 60101037004 &  ~47.84 &  26.48 &$30^{\prime\prime}$  & $30^{\prime\prime}$ \\ 
1ES 1101$-$232    & 2016-01-12   &  21:01:08 & 60101033002 & 101.87 & 50.79 & $40^{\prime\prime}$  & $40^{\prime\prime}$\\
1ES 1218$+$304    &  2015-11-23  &  01:06:08 & 60101034002 &  ~96.14 &  49.55 &$30^{\prime\prime}$  & $30^{\prime\prime}$ \\ 
H 2356$-$309      &  2016-05-18  &  16:31:08 & 60160840002 &  ~38.17 &  21.90 &$30^{\prime\prime}$  & $30^{\prime\prime}$\\

\hline

\end{tabular}

\end{table*}

\begin{table*}
\centering
\caption{X-ray variability parameters}
\label{tab:var_par}
\begin{tabular}{llcccccc}
  \hline
  Blazar Name  &  Obs. ID       & \multicolumn{3}{c}{ $F_{var}(percent)$} & ACF(ks) & Bin-size(ks) \\
             &                 &  Soft (3-10 keV) & Hard (10-79 keV) & Total (3-79 keV) &		&	\\
                 
  \hline
1ES 0347$-$121     & 60101036002  & $    	   -	     $	& $        -         $ & $        -	  $	&   -           & 1.00 \\ 
1ES 0414$+$009     & 60101035002  & $    	   -         $  & $        -	     $ & $  5.72 \pm 3.73 $	&   - 		& 1.00 \\ 
RGB J0710$+$591	 & 60101037004  & $    	   -   	     $  & $        -         $ & $ 	  -   	  $	&   - 		& 1.00 \\ 
1ES 1101$-$232   & 60101033002  & $ 4.24   \pm  1.10  $	& $  9.08 \pm  2.81  $ & $ 3.94  \pm 0.99 $	&   - 		& 2.00 	\\
1ES 1218$+$304 	 & 60101034002  & $  7.30  \pm  1.84 $	& $  7.28 \pm   8.70 $ & $ 7.62  \pm 1.49 $	&   23.51 	& 1.50 \\ 
H 2356$-$309	 & 60160840002  & $        -	     $  & $          -        $ & $ 2.10 \pm   3.40 $ 	&   - 		& 1.00 \\ 
\hline

\end{tabular}

\end{table*}

\begin{table*}
\centering

\caption{Model fits to the NuSTAR spectra}
\label{tab:spectra}
\resizebox{\textwidth} {!}{
\hskip-3.0cm\begin{tabular}{lccccccccccc}
  \hline
  Blazar Name &$n_H^{(a)}$ &  Obs. ID & \multicolumn{2}{c}{Power Law} & \multicolumn{3}{c}{ Log-parabola ($E_{pivot}=10$ keV) }&Flux$_{3-79 keV}^{(2)}$ & F-test & p-value\\
 \cmidrule[0.05cm](r){4-5}\cmidrule[0.05cm](r){6-8}	&       &    	  & $\Gamma $ &  $ \chi^2/dof $($\chi^2_r$) & $\alpha$ & $\beta $ & $ \chi^2/dof $($\chi^2_r$) &  &&\\
                 
  \hline
1ES 0347-121 & 3.05   & 60101036002  & $ 2.37 \pm 0.06 $ & 154.74/169 (0.92)& $ 2.47 \pm 0.10 $ & $ 0.37 \pm 0.25  $ & 148.37/168 (0.88)& $ 0.68 \pm 0.03 $   &	~7.21 &	$ 7.96 \times 10^{-3}$\\ 
1ES 0414+009 & 8.51   & 60101035002  & $ 2.77 \pm 0.06 $ & 164.66/182 (0.90)& $ 2.82 \pm 0.10 $ & $ 0.16 \pm 0.25 $ & 163.59/181 (0.90)& $ 0.71 \pm 0.02 $  & ~1.18 & 0.27 \\ 
RGB J0710+591 & 4.44 & 60101037004  & $ 2.27 \pm 0.03 $ & 401.23/371 (1.08)& $ 2.34 \pm 0.05 $ & $ 0.35 \pm 0.13 $ & 380.84/370 (1.02)& $ 2.41 \pm 0.06 $ & 19.81 & $ 1.13 \times 10^{-5}$\\ 
		 
1ES 1101$-$232 & 5.60 & 60101033002 & $2.50 \pm 0.02 $ & 640.45/579 (1.11)& $ 2.59 \pm 0.03 $ & $ 0.35 \pm 0.08 $ & 584.09/578 (1.01)& $ 2.94 \pm 0.07 $ & 55.78 & $ 3.02 \times 10^{-13}$\\ 
1ES 1218+304 & 1.94& 60101034002  & $ 2.55 \pm 0.03 $ & 361.34/366 (0.99)& $ 2.67 \pm 0.06 $ & $ 0.43 \pm 0.15 $ & 336.76/365 (0.92) & $ 1.19 \pm 0.03 $  & 26.64 & $ 4.03 \times 10^{-7}$\\ 
H 2356-309 & 1.44 & 60160840002  & $ 2.18 \pm 0.03 $ & 349.67/357 (0.98)& $ 2.23 \pm 0.04 $ & $ 0.27 \pm 0.13 $ & 336.91/356 (0.95)& $ 2.81 \pm 0.06 $ & 13.48 & $ 2.78 \times 10^{-4}$\\ 

  \hline

   \end{tabular}
}
(1) galactic hydrogen column density in units of $ 10^{20}$ cm$^{-2} $ taken from \cite{2005A&A...440..775K}, \\
(2) 3$-$79 keV unabsorbed flux for best fitted model in units of $ 10^{-11}$ erg cm$^{-2}$ s$^{-1}$
\end{table*}

\begin{table*}
\centering
\caption{Test of temporal variation of HR} 
 \label{tab:hr_var}
 \begin{tabular}{lcccc}\hline
Source	        & DoF & $\chi^2$  &  $\chi^2_{99}$  \\\hline
1ES 0347$-$121  & 119 & 112.11 & 157.80   \\ 
1ES 0414$+$009  & 129 & 125.07 & 169.27  \\
RGB J0710$+$591 & ~96 & ~76.12 & 131.14 \\
1ES 1101$-$232  & 188 & 230.50 & 236.02  \\
1ES 1218$+$304  & 183 & 214.25 & 230.42 \\
H 2356$-$309    & ~81 & ~86.92 & 113.51  \\\hline
\end{tabular}

\end{table*}

\begin{figure*}
\centering
\includegraphics[width=19cm, height=10cm]{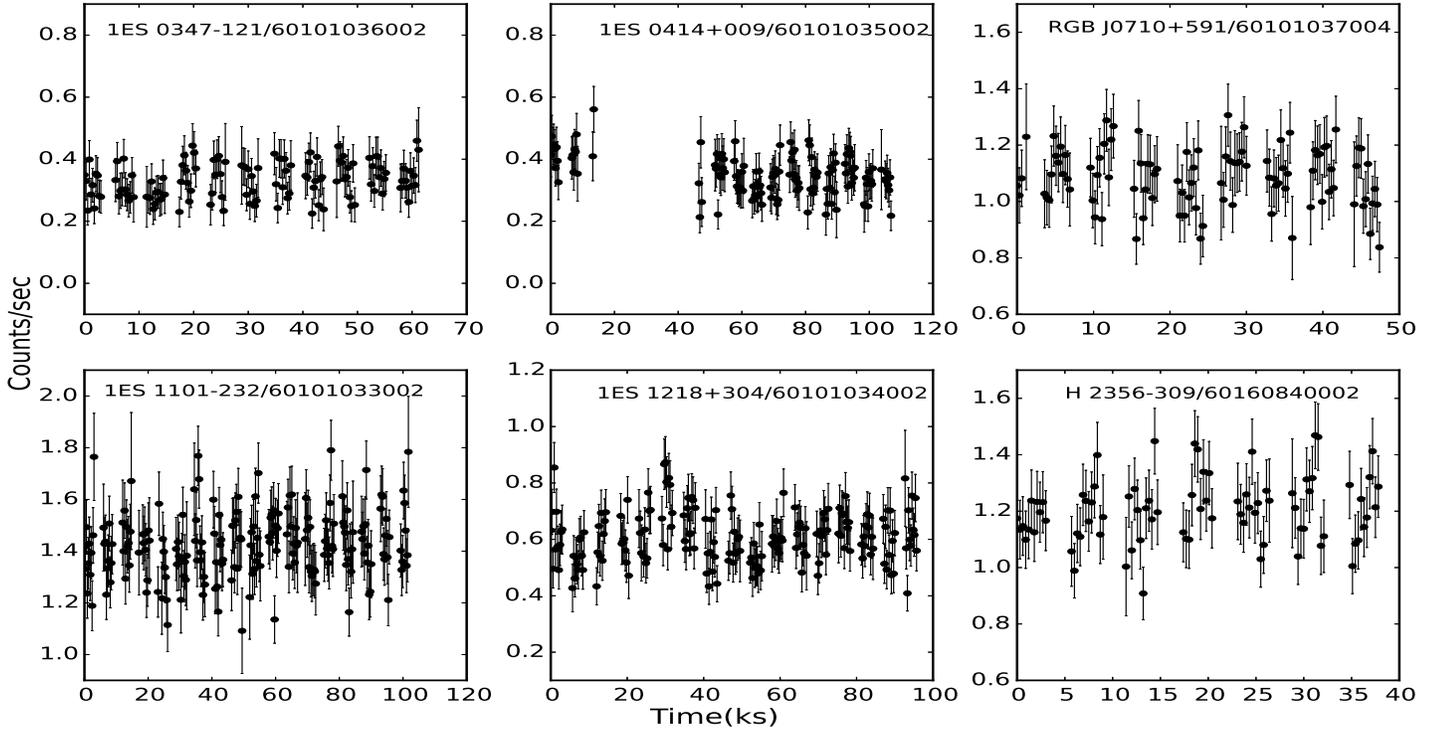}
\caption{\textit{NuSTAR} light curves of the TeV HBLs  1ES 0347$-$121, 1ES 0414$+$009, RGB J0710$+$591, 1ES 1101$-$232, 1ES 1218$+$304 and H 2356-309. The name of the blazar and the observation ID are given in each plot.} 
\label{fig:lc}
\end{figure*}

\begin{figure*}
\centering
\includegraphics[width=19cm, height=10cm]{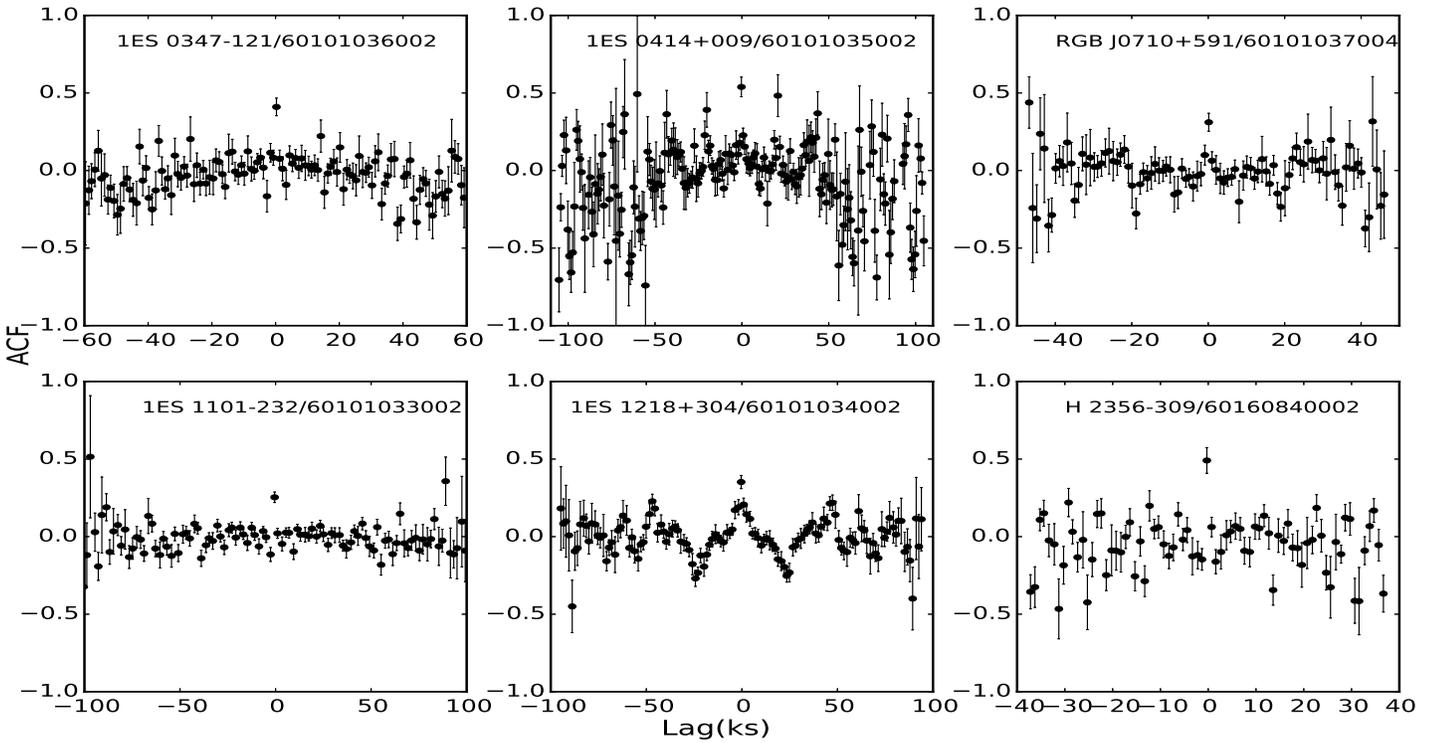}
\caption{Auto correlation plots for LCs of the TeV HBLs 1ES 0347$-$121, 1ES 0414$+$009, RGB J0710$+$591, 1ES 1101$-$232, 1ES 1218$+$304 and H 2356-309. The name of the blazar and the observation ID are given in each plot.} 
\label{fig:acf}
\end{figure*}


\section{ANALYSIS TECHNIQUES} \label{sec:analysis}
\subsection{Fractional Variance} \label{subsec:frac}
 
To estimate the amplitude of intraday variability in the LCs we used the fractional variance, which is commonly used for examing X-ray LCs (e.g. \cite{2002ApJ...568..610E,2003MNRAS.345.1271V,2016MNRAS.458.2350W}) and defined as (see \cite{2017ApJ...841..123P} for details)
\begin{equation}
F_{var} = \sqrt{\frac{S^2 - \overline{\sigma_{err}^2}}{{\bar{x}^2}}} .
\end{equation}
The uncertainty on $F_{var}$ is given by 
\begin{equation}
err(F_{var}) =  \sqrt{\left( \sqrt{\frac{1}{2N}}\frac{\overline{\sigma_{err}^2}}{\bar{x}^2 F_{var}} \right)  ^ 2+ \left(  \sqrt{\frac{\overline{\sigma_{err}^2}}{N}} \frac{1}{\bar{x}}\right) ^2 },
\end{equation}
where $S^2$ is the sample variance, $\bar{x}$ is the arithmetic mean of the LC, $\overline{\sigma_{err}^2}$ is the mean square error,  and $N$ is the total number of data points in the LC.  These values are given in Table \ref{tab:var_par}, where dashes indicate that the sample variances were smaller than the mean square errors so that no fractional variance could be claimed. We consider strong evidence for variability to be present when $F_{var} > 3 err(F_{var})$.

\subsection{Discrete Correlation Functions}
\label{sec:DCF}
We used a discrete correlation function (DCF) analysis, introduced by \citep{1988ApJ...333..646E},  to search for correlations between LCs in two energy bands. The way in which we use the DCF is explained in detail in \cite{2017ApJ...841..123P}. When the DCF is applied to the same LC it is called an auto correlation function, ACF, which can give any timescale of variability present in the LC.

\subsection{Hardness Ratio}
\label{sec:HR}
The hardness ratio (HR) is a crude method to examine spectral variations.  Given that earlier focusing X-ray telescopes have been restricted to exploring spectra only below 10 keV, we extracted LCs in two energy bands, here defining 3$-$10 keV as the soft band and 10$-$79 keV as the hard band.  We then computed a hardness ratio, which is defined as  
\begin{equation}
HR = \frac{(H-S)}{(H+S)},
\end{equation} 
 and the error in HR ($\sigma_{HR}$) is calculated as
\begin{equation}
\sigma_{HR}  = \frac{2}{(H+S)^2} \sqrt{(H^2 \sigma_S^2 + S^2 \sigma_H^2)},
\end{equation}
where $S$ and $H$ are the net count rates in the soft (3--10 keV) and hard (10--79 keV) bands, respectively, while $\sigma_S$ and $\sigma_H$ are their respective errors.

\subsection{Spectral Fitting}
\label{sec:SF}
Our  analysis of these \textit{NuSTAR} spectra was done with XSPEC\footnote{\url{https://heasarc.gsfc.nasa.gov/docs/xanadu/xspec/XspecManual.pdf}}  version 12.9.0. We grouped each spectra to a minimum of 20 counts per bin using FTOOL \textit{grppha} and then for each observation, the spectra of two \textit{NuSTAR} instruments, FPMA and FPMB, were  simultaneously fitted with two models via $\chi^2$ minimization. The first model we used for fitting is a power-law (PL):
\begin{equation}
F(E) = K E^{-\Gamma}
\end{equation}
where $\Gamma$ is the photon-index, $F(E)$ is the flux at energy $E$, and $K$ is the normalization parameter (photons keV$^{-1}$ cm$^{-2}$ s$^{-1}$).

The second model we applied is the log-parabola (LP) model. It is known that the X-ray spectra of many TeV HBLs are described well by the log-parabola model  \citep{2004A&A...413..489M,2007A&A...467..501T} defined as
\begin{equation}
F(E) = K (E/E_{pivot})^{-(\alpha+\beta \log(E/E_{pivot}))}
\end{equation}
where the free parameters $\alpha$, $\beta$ and K are the local photon index at fixed energy $E_{pivot}= 10$ keV, the spectral curvature, and the normalization parameter, respectively.  

The effect of galactic absorption was taken into account by multiplying each model with a \textit{phabs} component and taking fixed values of hydrogen column density, given in the second column of Table \ref{tab:spectra}.
The fitted model parameters for each model for all TeV HBLs are listed in Table \ref{tab:spectra}. The errors for each parameter are estimated to a 90\% confidence level ($\chi^2 = 2.706$). The model-fitted spectra together with data-to-model ratio for each of these six TeV HBLs are plotted in Fig.\ 4.

\section{RESULTS} \label{sec:result}

\subsection{1ES 0347$-$121} \label{subsec:0347}
The TeV HBL 1ES 0347$-$121 ($\alpha_{\rm 2000}$ = 03h49m23.0s; $\delta_{\rm 2000} = -11^{\circ}58^{\prime}38^{\prime\prime}$), at $z=0.188$ \citep{2005ApJ...631..762W}, was first detected in the Einstein Slew Survey \citep{1992ApJS...80..257E} and later classified as a BL Lac object \citep{1993ApJ...412..541S}. It was discovered as a very high energy (VHE) $\gamma-$ray emitter with an integral flux (at E $>$ 250 GeV) of 
$(3.9 \pm 1.1_{stat}) \times 10^{-12} {\rm cm}^{-2}{\rm s}^{-1}$ with HESS \citep{2007A&A...473L..25A}.

\textit{NuSTAR} observed 1ES 0347$-$121 on 2015 September 10 with a good exposure time of 32.93 ks. As seen from the LC, shown in Figure \ref{fig:lc}, the count rates are low and the data are noisy, so no detectable IDV is seen. We note that the \textit{NuSTAR} count rates for all these TeV blazars  are all less than 2 ct s$^{-1}$, whereas several of the set of 5 analyzed in Pandey et al.\ (2017) had means exceeding 5 ct s$^{-1}$.
Hence, the ACF plot, shown in Figure \ref{fig:acf}, is also noisy, providing no useful information.

The soft and hard LCs (left panel), HR plot (middle panel) and the DCF plot between soft and hard band (right panel) of 1ES 0347$-$121 are shown in Fig.\ 3. No significant spectral change is seen from the HR plot. We checked for variations quantitatively using a standard $\chi^2$ test, 
\begin{equation}
\chi^2 = \sum_{i=1}^{N} \frac{(x_i - \bar{x})^2}{\sigma_i^2},
\end{equation} 
\noindent where $x_i$ is the HR value, $\sigma_i$ is its corresponding error, and $\bar{x}$ is the mean HR value.
We considered a variation in the HR to be significant only if $\chi^2 > \chi^2_{99,\nu}$, where $\nu$ is the number of degrees of freedom (DoF) and the significance level is 0.99.    These results are given for all these blazars in Table \ref{tab:hr_var},  where we see that  for each source $\chi^2 < \chi^2_{99,\nu}$, so no significant spectral variations were detected.

The DCF plot is flat, which, in the presence of significant variations would indicate no correlation between the two energy bands, but since no variations are detectable this type of DCF is expected.

\subsection{1ES 0414$+$009} \label{subsec:0414}
The TeV blazar 1ES 0414$+$009 ($\alpha_{\rm 2000}$ = 04h16m53.0s; $\delta_{\rm 2000} = +01^{\circ}05^{\prime}20^{\prime\prime}$) is an HBL at $z=0.287$ \citep{1991AJ....101..818H}. It was first detected in X-rays with High Energy Astronomy Observatory (HEAO 1 A-1)\citep{1980ApJ...235..351U} and was classified as a BL Lac object by \cite{1983ApJ...270L...1U}. It was observed above 200 GeV by VERITAS with source flux equal to $(5.2 \pm 1.1_{stat} \pm 2.6_{sys}) \times 10^{-12} $photons cm$^{-2}$ s$^{-1}$ \citep{2012ApJ...755..118A}.

1ES 0414$+$009 was observed with \textit{NuSTAR} on 2015 November 25 with a good exposure time of 34.16 ks. The 3--79 keV LC of 1ES 0414$+$009 is shown in Figure \ref{fig:lc}. The data are both noisy and sparse, resulting in no significant IDV detection which is consistent with the value of fractional variance given in Table \ref{tab:var_par}. Consequently, the ACF plot, given in Figure \ref{fig:acf}, does not show any hint of variability timescale.

The soft and hard LCs of 1ES 0414$+$009 are plotted in the left panel of Fig.\ 3. The HR plot in the middle panel of that figure reveal no significant spectral variations  (also see Table \ref{tab:hr_var}) and the DCF plot shown in the right panel of Fig.\ 3 again is flat.

\subsection{RGB J0710$+$591} \label{subsec:0710}

The high-frequency-peaked BL Lacertae object RGB J0710$+$591 ($\alpha_{\rm 2000}$ = 07h10m26.4s; $\delta_{\rm 2000} = +59^{\circ}09^{\prime}00^{\prime\prime}$) is located at a redshift of $z=0.125$ \citep{1991ApJ...378...77G} and also was first detected by HEAO A-1 \citep{1984ApJS...56..507W}.  It was detected with VERITAS 
with the integral flux (above 300 GeV) recorded to be $(3.9 \pm 0.8 ) \times 10^{-12}$ cm$^{-2}$ s$^{-1}$\citep{2010ApJ...715L..49A}. 

\textit{NuSTAR} observed RGB J0710$+$591 on 2015 September 1 for 26.48 ks. As seen from the LC in Figure \ref{fig:lc} the source is somewhat brighter, by a factor of $\sim 3$ than the two discussed above, and there is a hint of variability. However, the data are still noisy and no significant intraday variations are found, as shown by the value of $F_{var}$, for which the error is two-thirds of the nominal value. The ACF of RGB J0710$+$591, shown in Figure \ref{fig:acf}, provides no evidence of an IDV timescale.

The soft and hard LCs (left panel), HR plot (middle panel) and the DCF plot of RGB J0710$+$591 are shown in Fig.\ 3. Given that both the soft and hard LCs are noisy, it is not surprising that the  HR plot does not show any detectable spectral variations (Table \ref{tab:hr_var}) and the DCF plot between the soft and hard band LCs is steady within the noise. 

\subsection{1ES 1101$-$232} \label{subsec:1101}
The TeV HBL 1ES 1101$-$232 ($\alpha_{\rm 2000}$ = 11h03m36.5s; $\delta_{\rm 2000} = -23^{\circ}29^{\prime}45^{\prime\prime}$), at $z=0.186$ \citep{1989ApJ...345..140R} was discovered in the Einstein Slew Survey \citep{1996ApJS..104..251P}.  \cite{2007A&A...470..475A} reported discovery of VHE $\gamma-$ray emission from 1ES 1101$-$232 with intergral flux (above 200 GeV) of $(4.5 \pm 1.2) \times 10^{−12} erg cm^{-2} s^{-1}$.

1ES 1101$-$232 was observed with \textit{NuSTAR} for 50.79 ks on 2016 January 12. This was the brightest of 
our blazars from this sample, at the time of its \textit{NuSTAR} observation and the LC in Figure \ref{fig:lc} appears to show significant flux variations in the energy range 3--79 keV. The value of $F_{var}$ for that full energy range given in Table \ref{tab:var_par} confirms the presence of IDV. The $F_{var}$ values for the soft and hard \textit{NuSTAR} bands, also given in Table \ref{tab:var_par}, confirm that the variations are present in both these  energy bands. Despite the presence of significant variability, the ACF plot of 1ES 1101$-$232 in Figure \ref{fig:acf} is almost flat, providing no variability timescale. 

The soft and hard LCs (left panel), HR (middle panel) and the DCF for 1ES 1101$-$232 are plotted in Fig.\ 3. The HR plot of 1ES 1101$-$232 reveals no detectable spectral variations, nor  is significant correlation observed from the DCF plot, despite the presence of variability.  In this case, these flat curves provide some evidence that the emission mechanism is the same for both bands. 

\subsection{1ES 1218$+$304} \label{subsec:1218}
1ES 1218$+$304 ($\alpha_{\rm 2000}$ = 12h21m26.3s; $\delta_{\rm 2000} = +30^{\circ}11^{\prime}29^{\prime\prime}$) is an HBL located at a redshift of $z=0.182$ \citep{2003A&A...412..399V}. \cite{2008ApJ...680L...9S} reported a flux over 2$-$10 keV range $\sim 2.0 \times 10^{-11}$ erg cm$^{-2}$ s$^{-1}$ and recently, an integrated flux of $3.33 \times 10^{-11}$ erg cm$^{-2}$ s$^{-1}$ in the 0.3$-$10 keV range was reported by \cite{2016MNRAS.458...56W}. The TeV flux (E $>$ 200 GeV) of  $(12.2 \pm 2.6) \times 10^{-12}$ cm$^{-2}$ s$^{-1}$ from 1ES 1218$+$304 was observed with VERITAS \citep{2009ApJ...695.1370A}.

\textit{NuSTAR} observed 1ES 1218$+$304 with a good exposure time of 49.55 ks on 2015 November 23. The LC and ACF plots of 1ES 1218$+$304 are shown in 
Figure \ref{fig:lc} and Figure \ref{fig:acf}, respectively. As seen from the LC, 1ES 1218$+$304 appears to show clear intraday variations which are confirmed by the $F_{var}$ value given in Table \ref{tab:var_par}. The variations are clearly present in the soft band but not obvious in the hard band, with its much lower fluxes. The ACF plot indicates a possible IDV timescale of $\sim$ 23.5 ks.

The soft and hard LCs of 1ES 1218$+$304 are plotted in the left panel of Fig.\ 3. The HR plot in the middle panel of that figure seems to show some fluctuations but is quite noisy, providing no useful information. The DCF plot shown in the right panel of Fig.\ 3 shows no correlations, as expected from the lack of significant variability in the hard band.

\subsection{H 2356$-$309} \label{subsec:2356}

H 2356$-$309 ($\alpha_{\rm 2000}$ = 23h56m09.4s; $\delta_{\rm 2000} = -30^{\circ}37^{\prime}23^{\prime\prime}$), located at a redshift of $z=0.165$ \citep{1991AJ....101..821F}, was first detected at X-rays by the Uhuru satellite \citep{1978ApJS...38..357F} and subsequently, by  HEAO A-1 \citep{1984ApJS...56..507W}. The X-ray (upto $\sim$ 50 keV) spectrum of H 2356$-$309 was characterized by a broken power-law with a synchrotron peak at $1.8 \pm 0.4$ keV during BeppoSAX observations  \citep{2001A&A...371..512C}. It was detected by H.E.S.S. with an integral flux (above 240 GeV) of $(3.06 \pm 0.26_{stat} \pm 0.61_{syst}) \times 10^{-12}$ cm$^{-2}$ s$^{-1} $\citep{2010A&A...516A..56H}.

H 2356$-$309 was observed with \textit{NuSTAR} for a relatively short good time exposure of 21.90 ks on 2016 May 18. As seen from the LC in the Figure \ref{fig:lc} the source had count rates nearly as high as 1ES 1101-232 and shows hints of variability.  However, the data are both noisier and sparser, and the $F_{var}$ value is consistent with no significant IDV. As a result, the ACF shown in Figure \ref{fig:acf} shows no IDV timescale. 

The soft and hard LCs (left panel), HR plot (middle panel) and the DCF plot (right panel) of H 2356$-$309 are shown in Fig.\ 3. The HR  plot reveal no significant spectral variation in the \textit{NuSTAR} range. The DCF plot is flat.

\begin{figure*}
\label{fig:lc_hr_dcf}
\centering
\includegraphics[width=19cm, height=20cm]{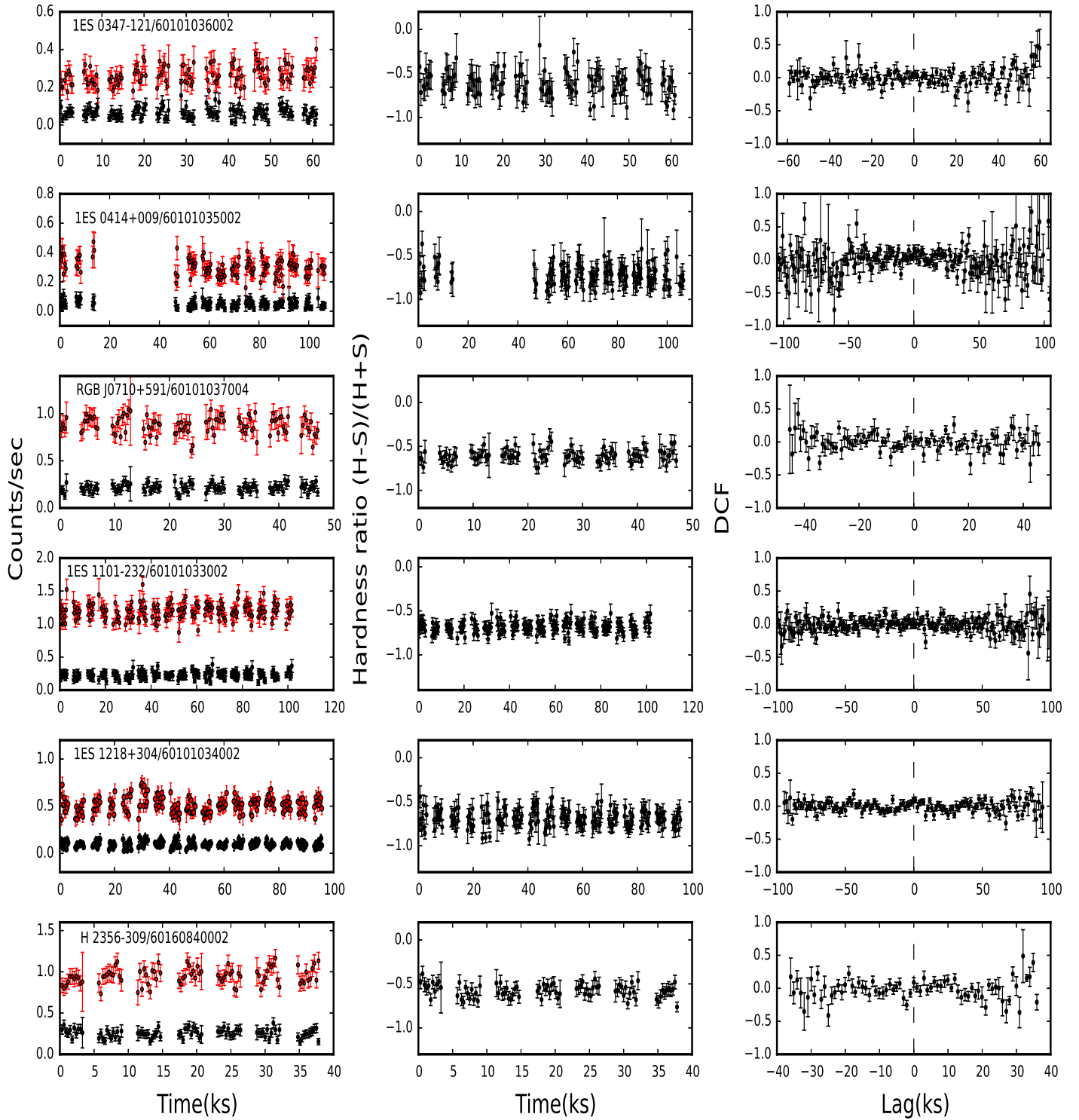}
\caption{\label{3_a} Soft (3-10 keV, denoted by red filled circles) and hard (10-79 keV, denoted by black filled circles) LCs (left panels), hardness ratios (middle panels), and the discrete correlation functions between soft and hard LCs (right panels) of the blazars 1ES 0347$-$121, 1ES 0414$+$009, RGB J0710$+$591,1ES 1101$-$232, 1ES 1218$+$304 and H 2356-309. The source names and observation ids are given in the left panels.}
\end{figure*}


\begin{figure*}[h]
\label{fig:spectra}
\centering
\begin{tabular}{ccc}
    \includegraphics[width=8cm,height=7.5cm]{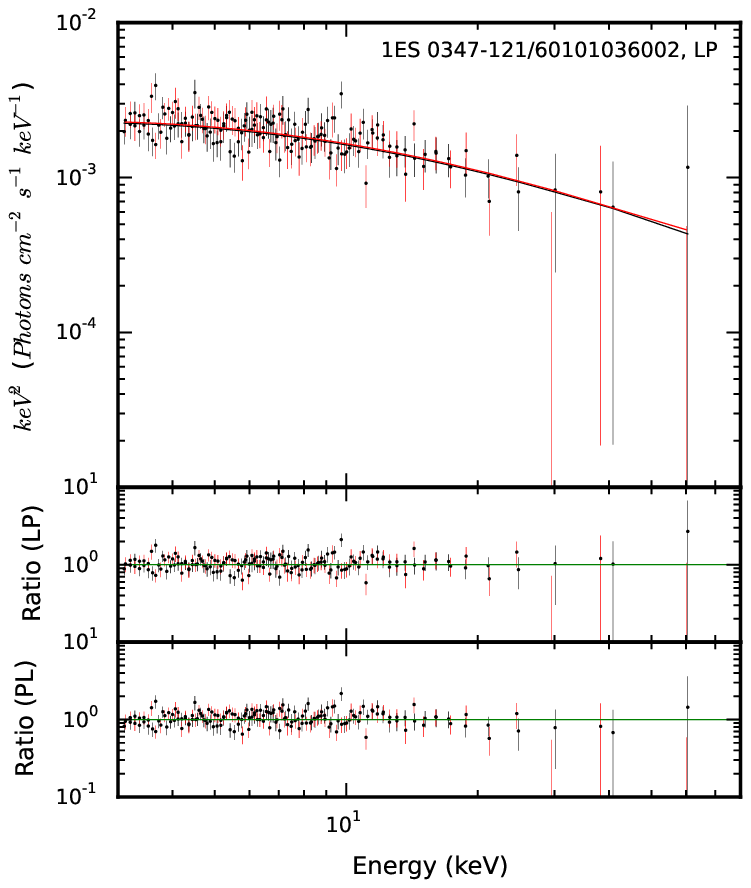}&
    \includegraphics[width=8cm,height=7.5cm]{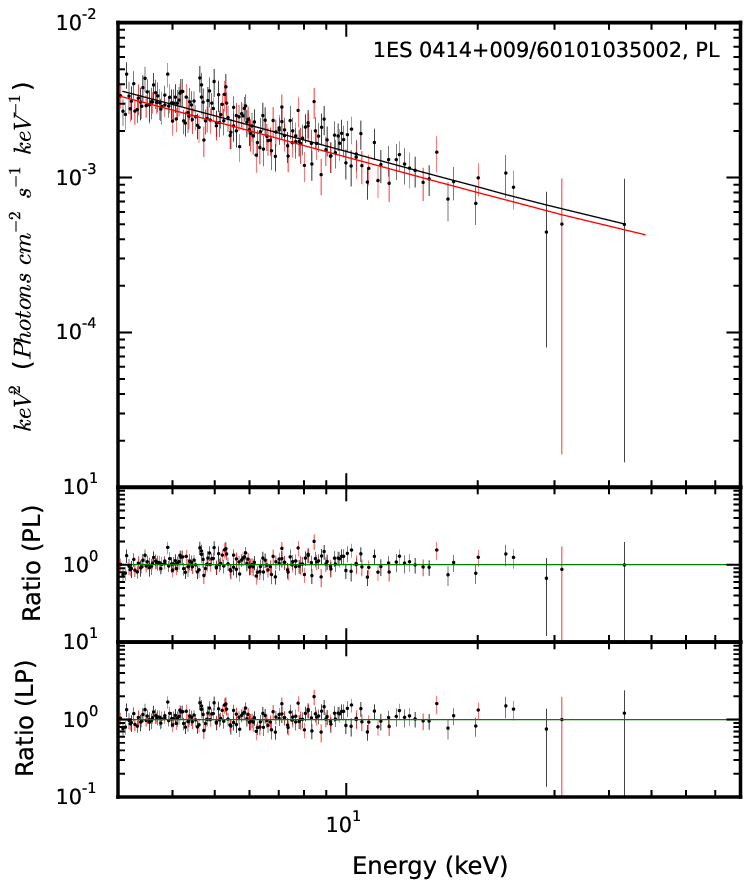}\\[2\tabcolsep]
    \includegraphics[width=8cm,height=7.5cm]{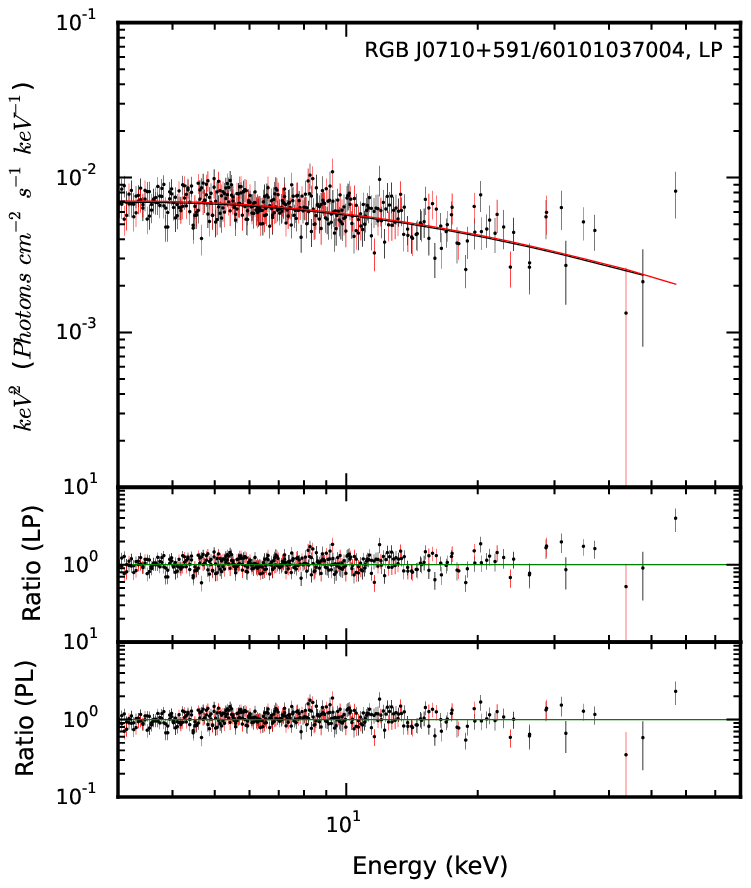}&
    \includegraphics[width=8cm,height=7.5cm]{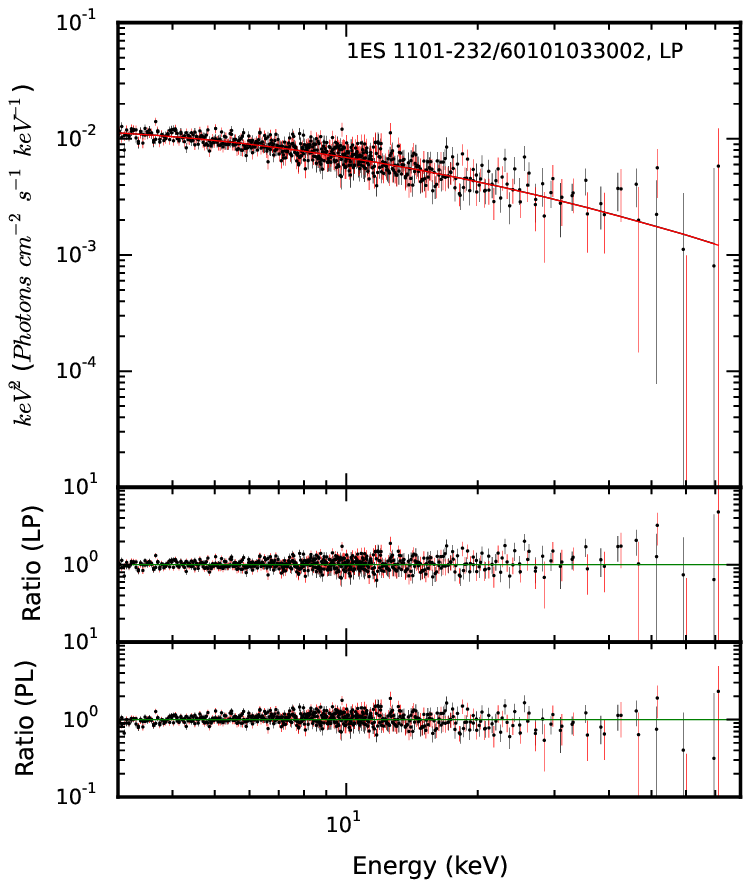}\\[2\tabcolsep]
    \includegraphics[width=8cm,height=7.5cm]{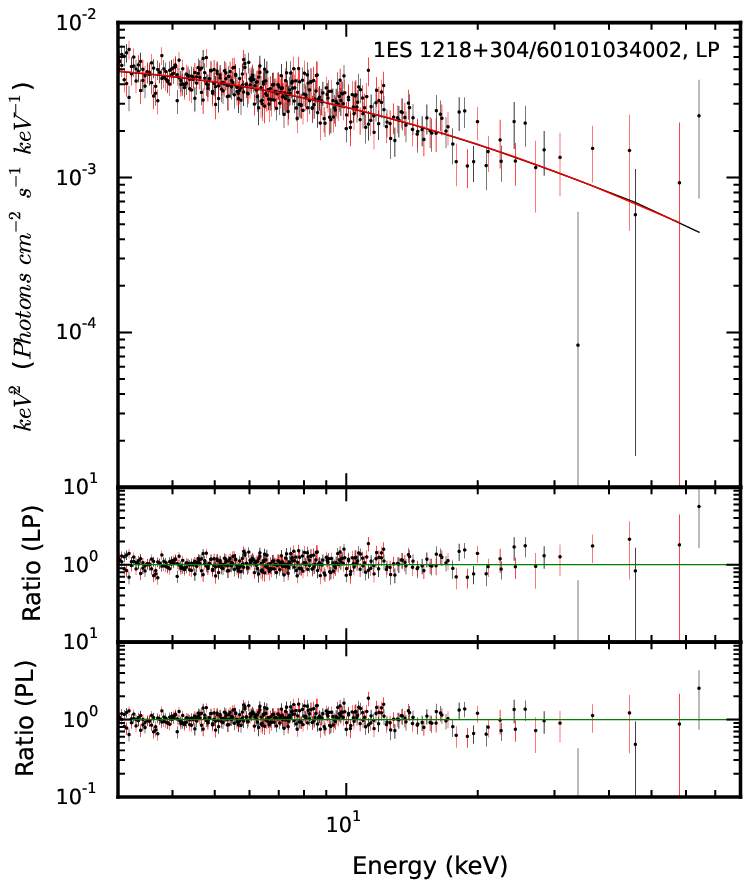}&
    \includegraphics[width=8cm,height=7.5cm]{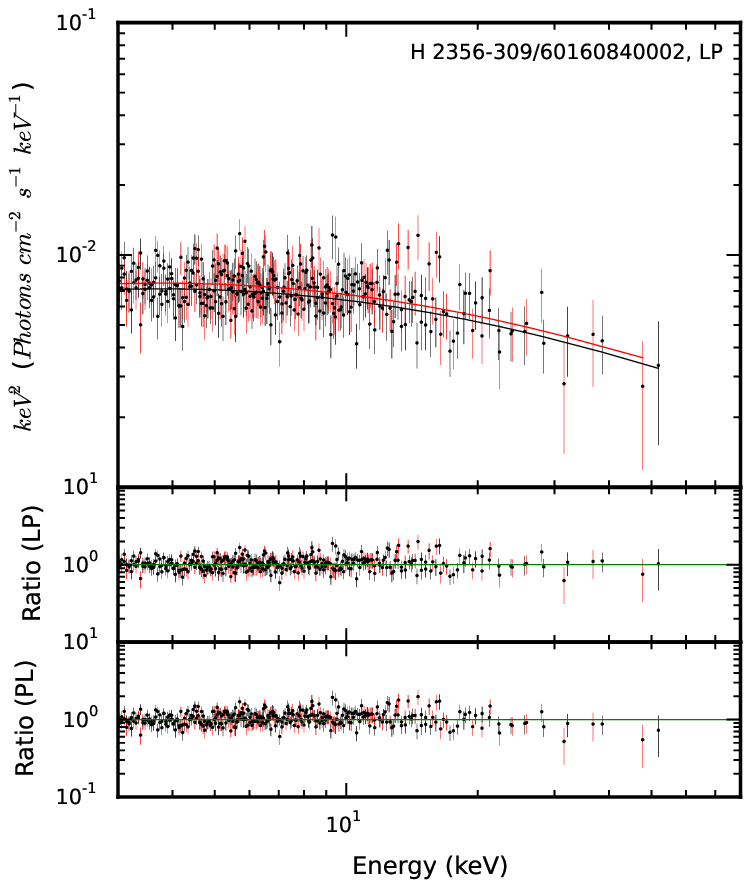}
\end{tabular}
\caption{\textit{NuSTAR} spectra (black points are for FPMA and red points are for FPMB) of six TeV HBLs with the best fitting model in the upper panels and the ratios (data/model)  for both the models tested, in the bottem two panels. The blazar name, observation ID, and the type best fitting model (LP or PL) are given in each plot.}
\end{figure*}

\section{DISCUSSION} 
\label{sec:discussion}
\subsection{Constraints on Physical Parameters from X-ray  Variability}
TeV blazars observed for sufficient times are known to exhibit strong variability with large amplitudes at all frequencies. Flux variations are understood to predominantly  originate from the Doppler-boosted relativistic jets \citep{2014ApJ...780...87M,2015JApA...36..255C}, however, in very low states, the instabilities or hot spots on the accretion disk can also produce variations on IDV and STV timescales \citep{1993ApJ...406..420M,1993ApJ...411..602C}.
At high energies (X-rays to $\gamma$-rays) the variations are often found to be very rapid, indicating compact emitting regions \citep{2004ApJ...605..662C, 2007ApJ...664L..71A, 2007ApJ...669..862A,2017ApJ...841..123P}.  The hard X-ray variability of $\sim 14 $ minutes detected in the \textit{NuSTAR} LCs of Mrk 421 \citep{2015ApJ...811..143P} were explained in terms of  magnetic reconnections accelerating particles to ultrarelativistic velocities in compact regions within relativistic jets  (``jets-in-a-jet" model; \cite{2009MNRAS.395L..29G}).

For these six, not extremely bright sources, we found a likely variability timescale for only one TeV HBL, 1ES 1218$+$304. The spectra clearly indicate that the dominant origin of hard X-rays in TeV HBLs is the high energy tail of the synchrotron emission. The synchrotron cooling timescale in the observer's frame is (e.g., \cite{2017ApJ...841..123P})
\begin{equation}
t_{cool}(\gamma)  \simeq 7.74 \times 10^8 \frac{(1+z)}{\delta}B^{-2}\gamma^{-1} ~{\rm s}, 
\end{equation}
where $\delta$ is the bulk Doppler factor, $B$ is the magnetic field strength in gauss (G) and $\gamma$ is the electron Lorentz factor.

In the \textit{NuSTAR} energy range the synchrotron frequency is (e.g., \cite{2017ApJ...841..123P})
\begin{equation}
\nu \equiv \nu_{19} \times10^{19}{\rm Hz} \simeq 4.2 \times 10^6 \frac{\delta}{1+z} B \gamma^2,
\end{equation}
where $0.08< \nu_{19} < 2$.
We eliminate $\gamma$ from  equations (6) and (7)  and use the fact that the cooling timescale must be smaller than or equal to the observed minimum variability timescale, to estimate for 1ES 1218$+$304 ($t_{var} = 23510$ s, $z = 0.182$ and $\delta \simeq 20$; e.g. \cite{2008ApJ...680L...9S}) that
\begin{equation}
B \geq 0.03~  \nu_{19}^{-1/3}  {\rm G}.
\end{equation}
Using equation (7) we can constrain the electron Lorentz factor to 
\begin{equation}
\gamma \leq 2.16 \times 10^6 \nu_{19}^{2/3}.
\end{equation}
For $\nu_{19} = 1 $,we get  $B \geq 0.03 $ G and $\gamma \leq 2.16 \times 10^6$. 

The characteristic size of the emitting region also can be estimated as 
\begin{equation}
R \leq c t_{var}  \delta /(1+z) \leq 1.19 \times 10^{16} ~{\rm cm}. 
\end{equation}
These values are close to those obtained by \cite{2008ApJ...680L...9S}. 

The maximum energy of photons produced by the electrons via Compton scattering (in the Thomson limit) can be estimated to
\begin{equation}
E_{max} \simeq \frac{\delta}{(1+z)} \gamma_{max} m_e c^2 \sim 19  \nu_{19}^{2/3}  ~{\rm TeV}. 
\end{equation}

\subsection{Correlation Between Emissions in Soft and Hard Energy Bands}

We searched for any possible correlation between the soft (3$-$10 keV) and hard (10$-$79 keV) band X-ray emissions for each TeV HBL using the discrete correlation function. The X-ray emissions in different energy bands are known to be generally well correlated \citep{2006ApJ...637..699Z,2017ApJ...841..123P}. However, in our analysis the DCF plots, shown in right panel of Fig.\ 3, are almost flat, indicating either of two possibilities.  The first is that the  X-ray emissions in these two energy bands are actually uncorrelated, which could indicate that the soft and hard X-ray emissions are plausibly  produced by different electron populations.  The second, and more likely, interpretation for the lack of correlations here is that the data are too noisy, particularly in the low flux,  hard X-ray LCs, to reveal the actual probable correlations between these bands for these sources. 

\subsection{X-ray Spectra}

We have calculated the unabsorbed 3--79 keV fluxes for each of these six  HBLs,  and they are given, with errors, in Table \ref{tab:spectra}.  They were determined using the \textit{cflux} routine of \textit{XSPEC} . The maximum flux observed is $\sim 2.94 \times 10^{-11}$ erg cm$^{-2}$ s$^{-1}$ for 1ES 1101$-$232 while the flux was least, at  $\sim 0.68 \times 10^{-11}$ erg cm$^{-2}$ s$^{-1}$, for 1ES 0347-121.

We performed simple hardness ratio analyses to search for spectral variations in the \textit{NuSTAR} energy range. In general, it has been found that for TeV HBLs the hardness ratio increases with increasing count rates, a behavior called 'harder when brighter' (e.g., \cite{2003A&A...402..929B,2004A&A...424..841R,2017ApJ...841..123P}). But in our current study we observed that the HR plots for all six TeV HBLs, shown in the middle panels of Fig.\ 3, do not show any significant variations. This indicates that during these observations we could not detect any variability of the X-ray spectra of these HBLs \citep{2008ApJ...682..789Z}. Again, this negative result could easily arise from the low count-rates for these blazars. 

We performed spectral fits using \textit{XSPEC} to study the  shape of these TeV blazar spectra in the 3--79 keV energy range, as displayed in Figure 4. We first applied the simple power-law (PL) model which has photon index ($\Gamma$) and the normalization as the two free parameters. As suggested by several studies that the X-ray spectra of HBLs are curved and described well with the log-parabola model (e.g. \cite{2004A&A...413..489M,2007A&A...466..521T,2007A&A...467..501T}), we also tried to fit the spectra with log-parabola (LP) model. The LP model has three free parameters: the photon index ($\alpha$) at a fixed energy ($E_{pivot}$),the curvature ($\beta$) and the normalization parameter. In this case the photon index is not a constant and varies along with the  logarithm of the energy (see eq.\ 5).

To examine any improvement in the fit using the LP over the PL model, we performed F-tests where the null hypothesis is that the simpler, PL, model provides the better fit. We found that for five out of six HBLs the curved log-parabola model provides a better fit over the simple power-law, as can be seen from the high F-statistic values and the corresponding probability ($0.99$) given in  Table \ref{tab:spectra}. Only in case of 1ES 0414$+$009 does a steep power-law with photon index $\simeq 2.77$ provide an equivalently good fit.
\textit{NuSTAR} spectra of the other five TeV BL Lac objects 1ES 0347$-$121, RGB J0710$+$591, 1ES 1101$-$232, 1ES 1218$+$304 and H 2356-309 are well described by LP models with $\alpha$ lying in the range 2.23--2.67 and the spectral curvature $\beta \simeq 0.27-0.43$.

The shape of X-ray spectra of TeV HBLs provides us with valuable information about the distribution of emitting particles and the particle acceleration mechanism. The curvature of X-ray spectra of BL Lac objects can be understood in terms of an energy-dependent particle acceleration probability and the subsequent  radiative cooling \citep{2004A&A...413..489M}. Thus, this study of the nature of X-ray spectra of TeV BL Lac objects may be used to understand the particle acceleration in these blazars.  Concave X-ray spectra of some TeV BL Lacs have  been reported in some studies (e.g. \cite{
2008ApJ...682..789Z}) and they were interpreted as a mixture of the high energy tail of the synchrotron emission and the low energy portion of the IC emission. However, we didn't find any signature of an IC component in our X-ray spectral analysis, indicating that the hard X-ray spectra of TeV HBLs are dominated by synchrotron emission.

\section{CONCLUSIONS} 
\label{conclusion}

We examined   the archival individual \textit{NuSTAR} LCs of  the six TeV HBLs that we had not previously analyzed \citep{2017ApJ...841..123P} for intraday variability and also searched for possible variability timescales using  discrete autocorrelation analyses. 
The X-ray count rates were quite low for these TeV HBLs and none of the exposure times exceeded 51 ks, so it should not be surprising that we found significant IDV only in the LCs of two of the six TeV HBLs, 1ES 1101$-$232 and 1ES 1218$+$304.

Using ACFs, we found a hint of the presence of a variability  timescale in the LC of only 1ES 1218$+$304. For the other 5 LCs, the ACF plots are noisy. Using that apparent observed variability timescale, we  estimated (for $\nu_{19} = 1 $) the magnetic field strength ($B \sim 0.03$ G), electron Lorentz factor ($\gamma \sim 2.2 \times 10^6 $) and emission region size ($R \sim 1.2 \times 10^{16} $ cm) for 1ES 1218$+$304.

 We  used a hardness ratio analysis to make a preliminary study of the X-ray spectral variability of these six TeV HBLs.  We found no significant variation in hardness ratio with time for each TeV HBL, indicating no detectable spectral variability was present during these observations. We also performed DCF analyses to search for any correlations between soft (3--10 keV) and hard (10--79 keV) \textit{NuSTAR} bands. The DCF plot for each TeV HBL is  almost flat, indicating either that  no significant correlation was present between the two energy bands, or, more likely, that the harder fluxes were too low to allow for any such correlations to be detected.
 
 The spectral shape of 1ES 0414$+$009 can be well fit with a 
 power-law. The \textit{NuSTAR} spectra of the remaining five HBLs, 1ES 0347$-$121, RGB J0710$+$591, 1ES 1101$-$232, 1ES 1218$+$304 and H 2356$-$309,  are clearly curved and require log-parabolic fits.

\vskip 0.25cm

\par This research has made use of data obtained with \textit{NuSTAR}, the first focusing hard X-ray mission managed by the Jet Propulsion Laboratory (JPL), and funded by the National Aeronautics and Space Administration (NASA). This research has also made use of the NuSTAR Data Analysis Software (NuSTARDAS) jointly developed by the ASI Science Data Center (ASDC, Italy) and the California Institute of Technology (Caltech, USA).

\bibliographystyle{aasjournal}
\bibliography{master}

\end{document}